\documentclass[conference,a4paper]{IEEEtran}

\IEEEoverridecommandlockouts

\usepackage{cite}
\usepackage{lmodern,textcomp}
\usepackage{graphics}
\usepackage{multicol}
\usepackage{lipsum}
\usepackage{color}
\usepackage{amsthm}
\usepackage{amssymb}
\usepackage{mathtools}
\usepackage{amsbsy}
\usepackage{setspace}
\usepackage{float}
\usepackage{lettrine}
\usepackage{newclude}
\usepackage[normalem]{ulem}
\usepackage{latexsym}
\usepackage{multirow}
\usepackage{rotating}
\usepackage{booktabs}
\usepackage{wasysym}
\usepackage{mathrsfs}
\usepackage{bbm} 
\usepackage{epsfig,cite,amsfonts,psfrag}
\usepackage{apptools}
\usepackage{graphicx}
\usepackage[table]{xcolor}
\usepackage{chngcntr}
\usepackage{blindtext}

\usepackage{hyperref}
\def\BibTeX{{\rm B\kern-.05em{\sc i\kern-.025em b}\kern-.08em
    T\kern-.1667em\lower.7ex\hbox{E}\kern-.125emX}}
    
\usepackage{optidef}
\usepackage{accents}
\usepackage{dsfont}
\usepackage{siunitx}
\usepackage{overpic}
\usepackage{amsmath,amssymb,amsfonts}
\usepackage{graphicx}
\usepackage{textcomp}
\usepackage{xcolor}
\usepackage{optidef}

\usepackage{graphicx}
\usepackage{cite}
\usepackage{subcaption}
\usepackage{caption}

\usepackage{lipsum}
\usepackage{overpic}
\usepackage{amssymb}
\usepackage{amsmath}
\usepackage{amsthm}
\usepackage{array}
\usepackage{color}
\usepackage{url}
 
\usepackage{optidef}
\usepackage{times}
\usepackage{fancyhdr,graphicx,amsmath,amssymb}
\usepackage[ruled,linesnumbered,noend]{algorithm2e}
\usepackage{algpseudocode}
\IEEEoverridecommandlockouts
\usepackage{cite}
\usepackage{graphicx}
\usepackage{url}
\usepackage{relsize}
\usepackage{hhline}
\usepackage{nicefrac}
\usepackage{amssymb,bm,upgreek}
\usepackage[table]{xcolor}
\usepackage{multirow}
\usepackage{subcaption}
\usepackage{amsthm}
\usepackage{caption}
\usepackage{xcolor}

\usepackage{optidef}
\usepackage{overpic}
\theoremstyle{remark}

\begin{document}

\title{Fronthaul Quantization-Aware MU-MIMO Precoding for Sum Rate Maximization
\thanks{This work was supported by the Knut and Alice Wallenberg Foundation.}
}

\author{\IEEEauthorblockN{Yasaman Khorsandmanesh, Emil Björnson, and Joakim Jaldén}
\IEEEauthorblockA{
\textit{KTH Royal Institute of Technology, Stockholm, Sweden}\\
Email: \{yasamank, emilbjo, jalden\}@kth.se}}

\maketitle

\begin{abstract}
This paper considers a multi-user multiple-input multiple-output (MU-MIMO) system where the precoding matrix is selected in a baseband unit (BBU) and then sent over a digital fronthaul to the transmitting antenna array. The fronthaul has a limited bit resolution with a known quantization behavior. We formulate a new sum rate maximization problem where the precoding matrix elements must comply with the quantizer. We solve this non-convex mixed-integer problem to local optimality by a novel iterative algorithm inspired by the classical weighted minimum mean square error (WMMSE) approach. The precoding optimization subproblem becomes an integer least-squares problem, which we solve with a new algorithm using a sphere decoding (SD) approach. We show numerically that the proposed precoding technique vastly outperforms the baseline of optimizing an infinite-resolution precoder and then quantizing it. We also develop a heuristic quantization-aware precoding that outperforms the baseline while having comparable complexity.
\end{abstract}

\begin{IEEEkeywords}
Sum rate maximization, weighted minimum mean square error, quantization-aware precoding.
\end{IEEEkeywords}

\section{Introduction}

Multi-user multiple-input multiple-output (MU-MIMO) systems enable high data rates through spatial multiplexing of multiple user equipments (UEs) on the same time-frequency resource \cite{Gesbert2007a}. A base station (BS) equipped with multiple antennas and channel state information (CSI) can transmit simultaneously to several UEs using different beamforming directivity to increase the sum rate, which is controlled by the precoding method. In precoding matrix design, it is most common to maximize the sum rate, or even the weighted sum rate, under a constraint on the total transmit power \cite{bjornson2012robust}; however, sum rate maximization is known to be NP-hard \cite{liu2010coordinated}. One popular approach for sum-rate maximization is the iterative weighted minimum mean square error (WMMSE) algorithm, which finds locally optimal solutions with affordable computational complexity \cite{shi2011iteratively}. This paper utilizes a novel iterative algorithm inspired by WMMSE.

A 5G BS typically consists of two main components: an advanced antenna system (AAS) and a baseband unit (BBU). The AAS is a box containing the antenna elements and their respective radio units (RUs). The BBU performs the digital processing related to the received uplink data and transmitted downlink data. The AAS and BBU are connected through a digital fronthaul. The integration of antennas and radios into a single box has made massive MU-MIMO practically feasible \cite{Bjornson2019d} and enabled the BBU to be virtualized in an edge cloud through migration to the centralized radio access network architecture \cite{peng2015fronthaul}. The new implementation bottleneck is the limited fronthaul capacity and the quantization errors it creates.

Both the uplink/downlink data and combining/precoding coefficients are sent over this digital fronthaul and must be quantized to a finite resolution. This paper proposes a novel linear block-level quantization-aware precoding technique that maximizes the sum rate.

\subsection{Prior Work}

Downlink MU-MIMO systems have been widely studied in previous literature regarding impairments in analog hardware \cite{Bjornson2014a} and the effect of low-resolution digital-to-analog converters \cite{jacobsson2017quantized}. These prior works are characterized by distortion created either in the RU, analog domain, or converters. Therefore, the transmitted signal is distorted after the precoding. The effect of limited fronthaul capacity is studied in \cite{parida2018downlink}, but the precoding design was not quantized. In \cite{khorsandmanesh2022optimized}, the authors proposed a fronthaul quantization-aware precoding design that minimizes the sum MSE, which will generally not maximize the sum rate. Many previous studies suggest designing a precoding matrix by maximizing the sum rate, often using the WMMSE approach; see \cite{christensen2008weighted,zhao2022rethinking} and references therein. Nevertheless, they consider ideal hardware or other types of distortion than precoding quantization.

\subsection{Contributions}

This paper proposes a transmit precoding design that finds a local optimum to the sum-rate maximization problem subject to a transmit power constraint over a limited-capacity fronthaul connection that only accepts precoding matrix elements from a discrete quantization codebook. 
The main contributions are:

\begin{itemize}
    \item We propose maximizing the sum rate by solving a quantization-aware precoding problem. As this mixed-integer problem is non-convex, we rewrite it following the iterative WMMSE algorithm to find a local optimum. Each iteration of the proposed iterative algorithm contains a new integer least-square problem that minimizes the weighted MSE at each UE. The solution is obtained in a new way inspired by sphere decoding (SD). We consider a reduced-complexity variation on the proposed algorithm where only the last iteration is quantization-aware. 
    
    \item We define quantization-unaware precoding as a baseline and then recommend a low-complexity heuristic algorithm to sequentially refine the quantization-unaware precoding columns to improve the sum rate. The complexity is comparable to quantization-unaware precoding; thus, massive MIMO scenarios can be effectively handled.
    
    \item We provide numerical results to compare different quantization-aware algorithms with quantization-unaware precoding baseline in terms of sum rate with the correlated Rician fading and a uniform planar array (UPA).
    
\end{itemize}

\section{System model}

We consider a single-cell  MU-MIMO downlink system, where the BS contains an AAS with $M$ antenna-integrated radios and serves $K$ single-antenna UEs. The AAS is connected to a BBU through a limited-capacity fronthaul link, which is modeled as a finite-resolution quantizer. The precoding matrix $\boldsymbol{P}$ is computed, and the data symbol vector $\boldsymbol{s}$  is encoded at the BBU and then sent over the fronthaul to the AAS. As data symbols are bit sequences from a channel code, we can transmit them over the fronthaul without quantization errors as they are already quantized. We can then map the data symbols obtained from the BBU to modulation symbols at the AAS. However, the precoding matrix computed at the BBU based on CSI is quantized due to the digital fronthaul. The quantized precoding matrix is then multiplied with the UEs' data symbols at the AAS, and finally, the product is transmitted wirelessly.

Before analyzing the proposed quantization-aware precoding in Section~\ref{sec3}, we introduce the problem formulation and quantization scheme in the following subsections.
\subsection{Problem Formulation}

The BBU uses its available CSI to select the downlink precoding matrix $\boldsymbol{P}$. As the main focus of this work is on managing quantization errors in a precoding matrix that maximizes the sum rate, we consider perfect CSI. However, the same algorithms can be used if the BBU has imperfect CSI but treats it as perfect. We postpone the channel estimation part for future work. The transmitted data symbol to the UE $k$ is denoted by $s_k$, which has zero mean and normalized unit power, and the corresponding channel vector  $\boldsymbol{h}^\mathrm{T}_k \in \mathbb{C}^{1 \times M}$ represents a narrowband channel 
and might be one subcarrier of a multi-carrier system. The algorithm developed in this paper can be applied individually to each subcarrier. The received signal at the UE $k$ is
\begin{equation}
    y_k = \boldsymbol{h}^\mathrm{T}_k {\boldsymbol{p}}_k s_k + \sum_{i=1,i\ne k}^K \boldsymbol{h}^\mathrm{T}_k {\boldsymbol{p}}_i s_i + n_k,
\end{equation}
where ${\boldsymbol{p}}_k\in \mathcal{P}^{M}$ is the quantized linear precoding vector for UE $k$ and $n_k \sim \mathcal{CN}(0,N_0)$ represents the independent additive complex Gaussian receiver noise with power $N_0$.
For later use, we define the total received signal as $\boldsymbol{y} = [y_1,\ldots,y_K]^\mathrm{T}$, the data symbols vector as $\textbf{s} = [s_1, \ldots, s_K]^\mathrm{T}\in \mathcal{O}^{K}$ ($\mathcal{O}$ is the finite set of constellation points such as a QAM alphabet), the channel matrix as $\boldsymbol{H} = [\boldsymbol{h}_1,\ldots,\boldsymbol{h}_K]^\mathrm{T} \in \mathbb{C}^{K\times M}$, and the precoding matrix as $\boldsymbol{P} = [\boldsymbol{p}_1,\ldots,\boldsymbol{p}_K] \in \mathcal{P}^{M\times K}$. 

The fronthaul quantization alphabet set $\mathcal{P}$ is defined as
\begin{equation}
   \mathcal{P} = \{ l_{R} + jl_{I} : l_{R},l_{I} \in \mathcal{L}  \}.
   \label{eq:quantizationset}
\end{equation}
We assume the same quantization alphabet is used for the real and imaginary parts. Here $\mathcal{L}= \{ l_0, \ldots ,l_{L-1} \}$ contains the set of real-valued quantization labels, $L=|\mathcal{L}|$ denotes the number of quantization levels, and 
$\Bar{L}=\log_2 (L)$ is the number of quantization bits per real dimension. Note that $\mathcal{P}$ becomes the complex-number set $\mathbb{C}$ in the case of infinite resolution. The quantized precoding matrix $\boldsymbol{P}$ and the data symbols vector $\boldsymbol{s}$ are sent separately over the fronthaul. The precoded signal vector $\boldsymbol{x}$ is calculated at the AAS as $ \boldsymbol{x} = \alpha \boldsymbol{P} \boldsymbol{s}$, where the scaling factor $\alpha = \sqrt{q/ \mathrm{tr}(\boldsymbol{P}\boldsymbol{P}^\mathrm{H})}$ is computed at the AAS, $q$ denotes the maximum transmit power, and $\mathrm{tr}( \cdot )$ denotes the matrix trace.
This scaling factor ensures that the  $\mathbb{E}[\left \Vert \textbf{x} \right \|_2^2] = q$, where $\| \cdot \|$ is the Euclidean norm so that the maximum power is always utilized, despite the finite-resolution quantization.

The achievable rate is $\log_2 (1 + \mathrm{SINR}_k (\boldsymbol{P}))$, where
the signal-to-interference-plus-noise-ratio (SINR) depends on the precoding matrix $\boldsymbol{P}$ as
\begin{equation}
    \mathrm{SINR}_k(\boldsymbol{P}) = \frac{\big| \boldsymbol{h}^\mathrm{T}_k {\boldsymbol{p}}_k \big|^2}{ \sum_{i=1,i\ne k}^K \big| \boldsymbol{h}^\mathrm{T}_k {\boldsymbol{p}}_i \big|^2 + N_0}.
\end{equation}
We want to maximize the sum rate of this downlink channel under the mentioned maximum transmit power constraint, and we define this problem as

\begin{maxi!}[2]
	  {\boldsymbol{P}\in \mathcal{P}^{M \times K} }{\sum_{k=1}^K  \log_2 \Big(1 + \mathrm{SINR}_k (\boldsymbol{P})\Big)\label{eq:WSR}}{\label{probelm1}}{\mathbb{P}_{1}: \ \ } \addConstraint{\mathrm{tr}(\boldsymbol{P}\boldsymbol{P}^\mathrm{H}) \le q,} \label{eq:power constrain}
\end{maxi!}
where the optimization variable is $\boldsymbol{P}$ with elements  $p_{m,k} \in \mathcal{P}$ for $k=1, \ldots ,K$ and $m=1, \ldots ,M$. 
Problem $\mathbb{P}_{1}$ is not convex since the utility is non-concave and the search space is discrete, so it is hard to find the optimal solution \cite{christensen2008weighted}.

\subsection{Quantization Scheme}
We are modeling the digital fronthaul as a quantizer. In practice, uniform quantization is often used, so we assume our quantizer function $\mathcal{Q}(\cdot) : \mathbb{C} \to \mathcal{P}$ is a symmetric uniform quantization with step size $\Delta$. Each entry of the quantization labels $\mathcal{L}$ is defined as 
\begin{equation}
    l_z  = \Delta  \left( z- \frac{L-1}{2}  \right), \quad z=0, \ldots, L-1.
\end{equation}
Furthermore, we let $\mathcal{T} = \{ \tau_0, \ldots, \tau_L \}$, where $-\infty = \tau_0 < \tau_1 < \ldots < \tau_{(L-1)} < \tau_{L} =\infty$, specify the set of the $L + 1$ quantization thresholds. For uniform quantizers, the thresholds
are 
\begin{equation}
    \tau_z =  \Delta  \left( z- \frac{L}{2}  \right), \quad z=1, \ldots, L-1.
\end{equation}
The quantizer function $\mathcal{Q}(\cdot)$ can be uniquely described by the set of quantization labels $\mathcal{L} = \{ l_z : z=0, \ldots, L-1 \}$ and the set of quantization thresholds $\mathcal{T}$. The quantizer maps an input $r \in \mathbb{C} $ to the quantized output $\mathcal{Q}(r) = l_o + jl_l \in \mathcal{P}$, where the set is defined in \eqref{eq:quantizationset}, if
$\mathfrak{R}\{ \mathcal{Q}(r) \} \in  [\tau_o,\tau_{o+1})$ and $\mathfrak{I}\{ \mathcal{Q}(r) \} \in  [\tau_l,\tau_{l+1})$.
The step size $\Delta$ of the quantizer should be chosen to minimize the distortion between the quantized output and the unquantized input. The optimal step size $\Delta$ depends on the dynamic range of the input, which in our case depends on the precoding scheme and channel model. 
We select the step size to minimize the distortion under the maximum-entropy assumption that each input element to the quantizer is  distributed $\mathcal{CN}(0,\frac{q}{KM})$, where the variance is selected so that the sum power of the elements matches with the power constraint \eqref{eq:power constrain}. The corresponding optimal step size for the normal distribution was numerically found in \cite{hui2001unifquantized}.

\section{Proposed WMMSE Algorithm}\label{sec3}

As the optimization problem $\mathbb{P}_1$ is non-convex, the global optimal solution is challenging to find. We instead target finding a local optimum. Inspired by \cite{shi2011iteratively}, we will rewrite $\mathbb{P}_1$ as an equivalent iterative WMMSE problem for which a local optimum can be found through alternating optimization. In the following, we decompose this equivalent optimization problem into a sequence of convex subproblems.

Let $\hat{s}_k = \beta_{k}{y}_k$ denote the estimate at UE $k$ of the transmitted data symbol $s_k$. It is obtained from the received signal $y_k$ using the receiver gain ${\beta}_k \in \mathbb{C}$ (also known as the precoding factor \cite{jacobsson2017quantized}). For a given receiver gain, the MSE in the data detection as  UE $k$ becomes
\begin{align}
    e_k(\boldsymbol{P}, {\beta}_k) &=  \mathbb{E} \left[   | {s}_k -  \hat{{s}}_k  |^2 \right]    \nonumber  
    \\ & =
    \left|\beta_{k}\right|^2\left(\left|\boldsymbol{h}^\mathrm{T}_k {\boldsymbol{p}}_k\right|^2+\sum\limits_{i=1,i\ne k}^K\left|\boldsymbol{h}^\mathrm{T}_k {\boldsymbol{p}}_i\right|^2+N_0\right) \nonumber \\& \quad -2\Re\left(\beta_{k}\boldsymbol{h}^\mathrm{T}_k {\boldsymbol{p}}_k \right)+1 \label{eq:mseUE}.
\end{align}
The MSE in \eqref{eq:mseUE} is a convex function of $\beta_k$. We can select the value of $\beta_k$ that minimizes the MSE for given $\boldsymbol{P}$ as
\begin{equation}
    \bar{\beta}_k (\boldsymbol{P})= \frac{ (\boldsymbol{h}^\mathrm{T}_k {\boldsymbol{p}}_k )^{*}}{\left|\boldsymbol{h}^\mathrm{T}_k {\boldsymbol{p}}_k\right|^2+\sum\limits_{i=1,i\ne k}^K\left|\boldsymbol{h}^\mathrm{T}_k {\boldsymbol{p}}_i\right|^2 + N_0}.
    \label{eq:beta}
\end{equation}
By plugging optimal $\bar{\beta}_k$ into \eqref{eq:mseUE}, we can see that $e_k(\boldsymbol{P}, \bar{\beta}_k)$ is equal to $1/(1+\mathrm{SINR}_k(\boldsymbol{P}) )$.

Now by defining the auxiliary weight $d_k \geq 0$, we
formulate a weighted sum MMSE problem subject to the same total transmit power constraint as in \eqref{eq:power constrain}:

\begin{mini!}[2]
  {\boldsymbol{P}\in \mathcal{P}^{M \times K} , \boldsymbol{\beta},\boldsymbol{d} }{\sum_{k=1}^K  \Big( d_k e_k(\boldsymbol{P}, {\beta}_k) - \log_2 (d_k)\Big) \label{eq:wmmse}}{\label{probelm2}}{\mathbb{P}_{2}: \ } \addConstraint{\mathrm{tr}(\boldsymbol{P}\boldsymbol{P}^\mathrm{H}) \le q,} 
\end{mini!}
where $\boldsymbol{\beta} = [\beta_1,\ldots,\beta_K]^\mathrm{T}$ is a vector containing all receiver gains and $\boldsymbol{d} = [d_1,\ldots,d_K]^\mathrm{T}$ is a vector containing all the UE weights in the weighted MSE. The problem $\mathbb{P}_2$ is equivalent to $\mathbb{P}_1$  in the sense that the optimal $\boldsymbol{P}$ is the same for both problems. This equivalence comes from the fact that the optimal weight for the UE $k$ is 
\begin{equation}
   \bar{d}_k = \frac{1}{e_k(\boldsymbol{P}, \bar{\beta}_k (\boldsymbol{P})) } = 1+\mathrm{SINR}_k(\boldsymbol{P}), \label{eq:weightsmse}
\end{equation}
so \eqref{eq:wmmse} then becomes $K-\sum_{k=1}^{K} \log_2(1+\mathrm{SINR}_k(\boldsymbol{P}))$.

The cost function in \eqref{eq:wmmse} is convex in each individual optimization variable, which is the key reason for considering this equivalent problem formulation.

\begin{figure}[t!]
  \centering
   \begin{overpic}[width=0.65\linewidth]{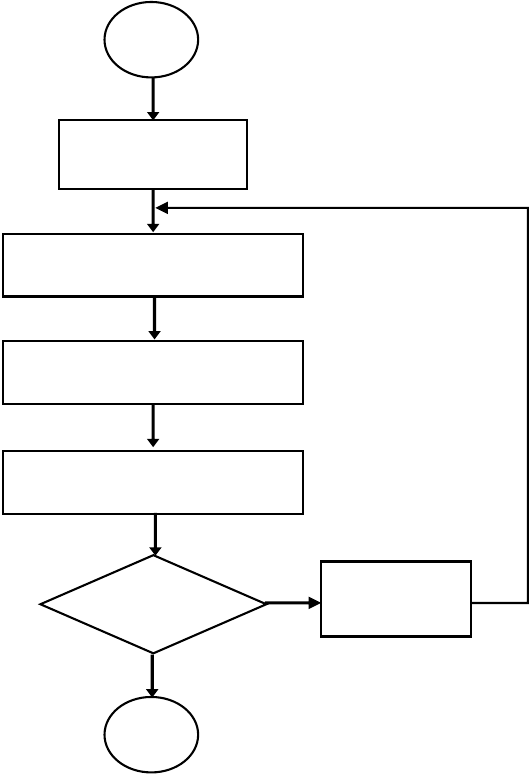}
   \put(15.8,94){Start}%
   \put(9,82){\scriptsize Initialize arbitrary}
   \put(10,79.5){\scriptsize$\boldsymbol{P}$, iteration $N$ }%
   \put(10,76.5){\scriptsize and set $n=1$}
   \put(3,65){\scriptsize Set $\beta_{k}$ \eqref{eq:beta}, $k=1,\ldots,K$}%
   \put(3,51){\scriptsize Set $d_k$  \eqref{eq:weightsmse}, $k=1,\ldots,K$}%
\put(7,37){\scriptsize Solve the  Problem $\mathbb{P}_3$}
   \put(17,23.5){\scriptsize $n $ is }
    \put(13,20){\scriptsize equal to $N$?}
    \put(35.5,23.5){\scriptsize No}
    \put(22,12){\scriptsize Yes }
   \put(42,24){\scriptsize Update $\boldsymbol{P}$ and}
   \put(42,20){\scriptsize set  $n = n+1$}
   \put(16.2,4){End}%
   \end{overpic} 
\caption{Flowchart of the proposed iterative algorithm for solving WMMSE problem $\mathbb{P}_2$. Here $N$ is the maximum number of iterations.}
\vspace{-5mm}
\label{fig:algorithm}
\end{figure}
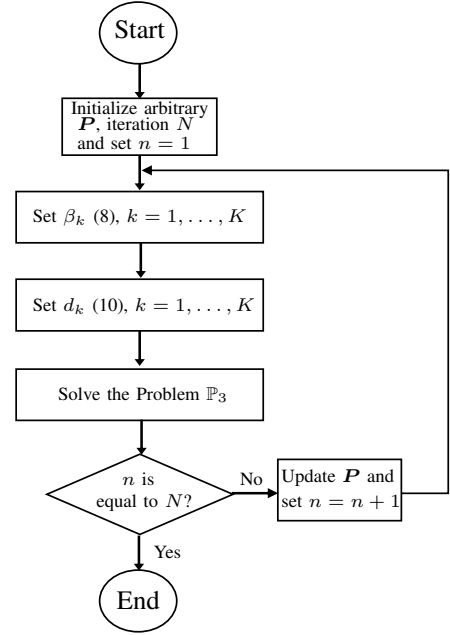

For fixed ${\beta}_{k}$  and ${d}_k$ (e.g., calculated as in \eqref{eq:beta} and  \eqref{eq:weightsmse}), we have the WMMSE problem
\begin{mini!}[2]
	  {\boldsymbol{P}\in \mathcal{P}^{M \times K}}{\sum_{k=1}^K   {d}_k e_k(\boldsymbol{P}, {\beta}_k) \label{eq:sum-SE-maximization-weighted-MMSE-subproblem}}{\label{probelm3}}{\mathbb{P}_{3}: \ } \addConstraint{\mathrm{tr}(\boldsymbol{P}\boldsymbol{P}^\mathrm{H}) \le q,} 
	  \label{eq:power constrainp3}
\end{mini!}
which is a mixed-integer convex problem. It can be solved using  general-purpose methods, such as  CVX \cite{grant2014cvx}.
By iterating between updating 
$\beta_k$ using \eqref{eq:beta}, $d_k$ using \eqref{eq:weightsmse}, and $\boldsymbol{P}$ by solving $\mathbb{P}_3$, we obtain a block coordinate descent algorithm that will converge to a stationary point (for the same reasons as in \cite{shi2011iteratively}). 
This algorithm is summarized in Figure~\ref{fig:algorithm}. 

We can initialize the algorithm using any precoding matrix, including those obtained using classical infinite-resolution precoding schemes.
The Wiener filtering (WF) precoding scheme (also known as regularized zero-forcing) is the most desirable one in this context since it can  be derived by minimizing the sum MSE \cite{joham2005linear}. Hence, we suggest setting the initial  precoding matrix as $\boldsymbol{P}_{\mathrm{initial}} = \boldsymbol{H}^\mathrm{H} (\boldsymbol{H}\boldsymbol{H}^\mathrm{H}+\frac{KN_0}{q}\textbf{I}_K )^{-1}$.

\subsection{Efficient Implementation of Quantization-Aware Precoding}

The main complexity in the proposed algorithm originates from solving $\mathbb{P}_3$. Instead of using a general-purpose solver, we will propose an efficient dedicated algorithm.
We can rewrite the objective function \eqref{eq:sum-SE-maximization-weighted-MMSE-subproblem} as 
\begin{align}
 & \sum_{k=1}^K {d}_{k}e_{k}\left(\boldsymbol{P},{\beta}_k\right) = \mathbb{E} \left[  \left \Vert \sqrt{\mathrm{diag}(
 {\boldsymbol{d}}
 )} \Big (    \boldsymbol{s} -   \mathrm{diag}(
 {\boldsymbol{\beta}}) \boldsymbol{y} \Big) \right \|_2^2 \right],
 \label{eq:mmseform}
\end{align} 
where $\mathrm{diag}(\boldsymbol{d})$ is a diagonal matrix with elements from the vector $\boldsymbol{d}$ with UE weights on the main diagonal. The expression in \eqref{eq:mmseform} can be expanded as 

\begin{align}
&\mathbb{E} \left[  \left \Vert \sqrt{\mathrm{diag}(\boldsymbol{d})}    \boldsymbol{s} -  \sqrt{\mathrm{diag}(\boldsymbol{d})} \mathrm{diag}(\boldsymbol{\beta}) \boldsymbol{y} \right \|_2^2 \right] \nonumber  
\\ &= \mathrm{tr} \Big(\mathrm{diag}(\boldsymbol{d})  - \sqrt{\mathrm{diag}(\boldsymbol{d})}\boldsymbol{D} \boldsymbol{H}\boldsymbol{P}-  \sqrt{\mathrm{diag}(\boldsymbol{d})} \boldsymbol{P}^\mathrm{H}
 \boldsymbol{H}^\mathrm{H}\boldsymbol{D}^\mathrm{H} \nonumber \\& \quad + \boldsymbol{D}\boldsymbol{H}\boldsymbol{P}\boldsymbol{P}^\mathrm{H}\boldsymbol{H}^\mathrm{H}\boldsymbol{D}^\mathrm{H}   +  N_0 \boldsymbol{D} \boldsymbol{D}^\mathrm{H}  \Big), \label{eq:simplified} 
\end{align} 
where we introduce the notation $\boldsymbol{D}= \sqrt{\mathrm{diag}(\boldsymbol{d})} \mathrm{diag}(\boldsymbol{\beta})$.

We first notice that $\mathbb{P}_3$ is a so-called integer least-squares problem due to \emph{uniform quantizer} that we chose. The search space is a scaled finite subset of the infinite integer lattice. A technique that has previously been proposed as an efficient algorithm to solve closest lattice point problems in the Euclidean sense is called sphere decoding (SD). SD has significantly lower average computational complexity than a naive exhaustive search \cite{jalden2005complexity}. The basic principle of SD is to reduce the number of search points in a skewed lattice that lies within a hypersphere of radius $d$, which can speed up the process of finding the solution without loss of optimality. In this paper, we consider the Schnorr-Euchner SD (SESD) algorithm \cite{agrell2002closest}, where the enumeration sorts candidate symbols in a zig-zag manner. SESD improves the basic SD algorithm by first checking the smallest child node of the parent node in each layer because the first found the feasible solution is often quite suitable and quickly enables a reduction of the search radius. 
Thus, many branches can be pruned, and the calculation complexity can be further lowered.

We propose using the SESD algorithm to approximately solve $\mathbb{P}_3$ for fixed values of $\beta_k$ and $d_k$. However, the classical SD framework
does not contain any power constraint. To adapt our problem $\mathbb{P}_3$ to match with the form required by SD algorithms, we need to proceed as follows. First, we rewrite the objective
function \eqref{eq:simplified} using the Lagrange multiplier $\lambda$ as 
\begin{align}
    \mathfrak{L}(\boldsymbol{P}, \boldsymbol \beta, \lambda) &= \mathrm{tr} \Big(\mathrm{diag}(\boldsymbol{d})  - \sqrt{\mathrm{diag}(\boldsymbol{d})}\boldsymbol{D} \boldsymbol{H}\boldsymbol{P}  \nonumber \\ & -  \sqrt{\mathrm{diag}(\boldsymbol{d})} \boldsymbol{P}^\mathrm{H}
 \boldsymbol{H}^\mathrm{H}\boldsymbol{D}^\mathrm{H}   + \boldsymbol{D}\boldsymbol{H}\boldsymbol{P}\boldsymbol{P}^\mathrm{H}\boldsymbol{H}^\mathrm{H}\boldsymbol{D}^\mathrm{H}  \nonumber \\& +  N_0 \boldsymbol{D} \boldsymbol{D}^\mathrm{H}  \Big) + \lambda \big( \mathrm{tr}(\boldsymbol{P}\boldsymbol{P}^\mathrm{H})-q \big). \label{eq:lagrangem}
\end{align}
When minimizing \eqref{eq:lagrangem} with respect to $\boldsymbol{P}$ and $\lambda$, we can drop the constant term $\mathrm{tr} (\mathrm{diag}(\boldsymbol{d}) )$ and have problem $\mathbb{P}_4$, given in \eqref{eq:lastsd} at the top of the next page. Although strong duality does not hold for the integer least-squares problem $\mathbb{P}_3$, $\mathbb{P}_4$ provides a good approximation. 
For solving $\mathbb{P}_4$, we can utilize the SD algorithm for a fixed value of $\lambda$. We then make a bisection search over $\lambda$ to find the value that gives a solution that satisfies the power constraint in \eqref{eq:power constrainp3} near equality.

For a fixed value of $\lambda$, and by vectorizing  $\mathbb{P}_4$ and using  $\boldsymbol{f} = \mathrm{vec}( (\sqrt{\mathrm{diag}(\boldsymbol{d})}\boldsymbol{DH})^\mathrm{T})$, we  obtain $\mathbb{P}_5$, 
\begin{figure*}[h!]
\begin{mini}|l|
	  {\boldsymbol{P}\in \mathcal{P}^{M \times K}, \lambda \ge 0  }{\mathrm{tr}\Big(\boldsymbol{P}^\mathrm{H}\big(\boldsymbol{H}^\mathrm{H}\boldsymbol{D}^\mathrm{H} \boldsymbol{D}\boldsymbol{H} + \lambda\boldsymbol{I}_M \big)\boldsymbol{P}- \sqrt{\mathrm{diag}(\boldsymbol{d})} \boldsymbol{D}\boldsymbol{H}\boldsymbol{P} - (\sqrt{\mathrm{diag}(\boldsymbol{d})} \boldsymbol{D}\boldsymbol{H}\boldsymbol{P})^\mathrm{H} \Big) 
	  - \lambda q .}{\label{eq:lastsd}}{\mathbb{P}_{4}: \ } 
\end{mini}
\end{figure*}
\begin{figure*}[h!]
\begin{mini}|l|
	  {\boldsymbol{p}_i\in {\mathcal{P}}^{M }, i=1,\ldots,K }{\sum_{i=1}^K \Big( \boldsymbol{p}_i^\mathrm{H} \left ( \boldsymbol{H}^\mathrm{H}\boldsymbol{D}^\mathrm{H} \boldsymbol{D}\boldsymbol{H}+\lambda \boldsymbol{I}_M \right )\boldsymbol{p}_i -\boldsymbol{f}_i^\mathrm{T}\boldsymbol{p}_i - \left ( \boldsymbol{f}_i^\mathrm{T}\boldsymbol{p}_i\right )^\mathrm{H}\Big),}{ \label{eq:modifysd}}{\mathbb{P}_{5}: \ }
\end{mini}
\hrule
\end{figure*}
which finds a suboptimal precoding matrix and has $K$ separable objective functions that each only depends on one of the optimization variables.
This feature enables \emph{parallel} optimization of $\boldsymbol{p}_i$ for $i=1,\ldots,K$. Thus, in addition to the more efficient search strategy, the reformulation of problem $\mathbb{P}_5$ also significantly reduces the dimension of each subproblem \cite{khorsandmanesh2022optimized}. By defining $\hat{\boldsymbol{V}} = \boldsymbol{H}^\mathrm{H}\boldsymbol{D}^\mathrm{H} \boldsymbol{D}\boldsymbol{H}+\lambda \boldsymbol{I}_M$, we can obtain the equivalent formulation of each term of the objective function in \eqref{eq:modifysd} as 
\begin{align}
    \boldsymbol{p}_i^\mathrm{H} \hat{\boldsymbol{V}} \boldsymbol{p}_i -\boldsymbol{f}_i^\mathrm{T}\boldsymbol{p}_i - \left ( \boldsymbol{f}_i^\mathrm{T}\boldsymbol{p}_i\right )^\mathrm{H} \label{eq:sd} = \lVert \boldsymbol{c}_i - \boldsymbol{G}\boldsymbol{p}_i \rVert_2^2 - \boldsymbol{c}_i^\mathrm{H}\boldsymbol{c}_i, 
\end{align}
where $\boldsymbol{G} \in \mathbb{C}^{M \times M }$ is obtained from the Cholesky decomposition $\hat{\boldsymbol{V}} = \boldsymbol{G}^\mathrm{H} \boldsymbol{G}$ and $\boldsymbol{c}_i = (\boldsymbol{f}_i^\mathrm{T}\boldsymbol{G}^{-1})^\mathrm{H}$. We can now minimize \eqref{eq:sd} by the classical SESD algorithm. We call this approach the \textit{Proposed SD-based WMMSE} algorithm.

\subsection{Quantization-Unaware Precoding}

The conventional WMMSE-based algorithms in \cite{christensen2008weighted,zhao2022rethinking} (among others) compute a precoding matrix from $\mathbb{C}^{M\times K}$ that maximizes the sum rate with infinite resolution and the available CSI. Hence, the naive baseline approach would be to compute such a  precoding matrix $\boldsymbol{P}_{\mathrm{unquantized}}=\boldsymbol{W} \in \mathbb{C}^{M\times K}$ and then quantize each entry using $\mathcal{Q}(\cdot): \mathbb{C}^{M \times K} \to \mathcal{P}^{M \times K}$ so that the result can be sent over the fronthaul. In this case, the BBU follows Fig.~\ref{fig:algorithm} but solves $\mathbb{P}_3$ in the domain $\mathbb{C}$ instead of $\mathcal{P}$, and $\mathbb{P}_3$ becomes a continuous convex optimization problem that is solvable using any general-purpose convex solver.
We will refer to this quantization-unaware precoding approach as the \emph{Unaware WMMSE}. After the WMMSE algorithm has converged, the final precoding matrix is obtained by quantizing each entry as $\boldsymbol{P} = \mathcal{Q}(\boldsymbol{W})$. 
\subsection{Combined Quantization-Aware and -Unaware Precoding}

Although the proposed SD approach is tailored for the problem, there is a limit to how large setups ($M$ and $K$) it can handle before the run time of the \textit{Proposed SD-based WMMSE} approach becomes an issue. As we run the SD algorithm in each iteration, the algorithm's average complexity over $N$ iterations becomes $O(NKL^{2\gamma M})$ for some $0 \le \gamma \le 1$ \cite{jalden2005complexity}.

An alternative would be to search for a precoding matrix in $\mathbb{C}^{M \times K}$ for $N-1$ iterations and then use the SD-based algorithm only for the final iteration. In this case, we will first identify UE gains $\beta_k$ and weights $d_k$ that are suitable for sum-rate maximization with infinite-resolution precoding and then compute the corresponding optimized quantization-aware precoding. 
The order of complexity would be  $O(KL^{2\gamma M})$ for some $0 \le \gamma \le 1$. We call this method \emph{Half-aware WMMSE}.

\subsection{Heuristic Quantization-Aware Precoding}

Although the \textit{Half-aware WMMSE} scheme has lower complexity than the \textit{Proposed SD-based WMMSE}, the complexity grows exponentially with the number of antennas $M$, which makes scenarios with  many antennas intractable. Hence, we believe it should primarily be seen as a benchmark for designing quantization-aware precoding schemes with low complexity. In this subsection, we propose such a heuristic precoding scheme. The proposed scheme is an add-on to the \textit{Unaware WMMSE} precoding. After we quantize the precoding matrix $\boldsymbol{W} \in \mathbb{C}^{M \times K}$ to obtain $\boldsymbol{P} \in \mathcal{P}^{M \times K}$, we refine the elements sequentially. We consider the
second closest quantization levels in both the real and imaginary dimensions  according to Euclidean distance \cite{khorsandmanesh2022optimized}. This search gives us three alternative ways of quantizing each element in the precoding matrix $\boldsymbol{W}$. We call this method the \emph{Heuristic quantization-aware}.

As the matrix elements are refined sequentially, we must order the elements properly. We propose to start by updating the column of the quantized precoding matrix corresponding to the UE $k$ with the highest \textit{generated interference} $\mathrm{GI}_k = \sum_{{i=1} , i \ne k}^K |[{\boldsymbol{H}} {\hat{\boldsymbol{P}}}]_{i,k}|^2$, where $\hat{\boldsymbol{P}} = \alpha \boldsymbol{P} = \alpha \mathcal{Q} (\boldsymbol{W})$ since this might improve the performance the most.\footnote{We have noticed experimentally that this leads to the largest improvement in sum rate at high SNR.} Then for that specific UE $k$, for each transmit antenna $m \in \{ 1, \ldots, M \}$, we identify the four nearest points in $\mathcal{P}$ to the element $w_{k,m}$ from the original unquantized precoding matrix $\boldsymbol{W}$.
We evaluate the sum rate
\begin{equation}  \sum_{k=1}^K 
\log_2 \left(1+\frac{\big|[{\boldsymbol{H}} \hat{\boldsymbol{P}}]_{k,k}\big|^2}{\sum_{i=1, i  \ne k}^{K}\big|[{\boldsymbol{H}} \hat{\boldsymbol{P}}
]_{k,i}\big|^2 +N_0} \right),
\label{eq:sumrate}
\end{equation}
for the four different ${\boldsymbol{P}}$ options obtained with $p_{k,m} \in \{ \text{four nearest points to } w_{k,m} \hspace{1mm} \text{in} \hspace{1mm} \mathcal{P} \} $ while all other elements are fixed. 
We then replace the corresponding element in $\boldsymbol{P}$ with the option that achieves the largest sum rate. The rest of the UEs are ordered based on decreasing generated interference, and the precoding elements are updated similarly.

\section{Numerical Results}\label{sec4}

This section compares the sum rates achieved by the aforementioned precoding approaches as a function of the SNR. The sum rate is calculated using Monte Carlo simulations for the case of Gaussian signaling. 

\subsection{Channel Model}
We consider spatially correlated Rician fading channels composed of a line-of-sight (LoS) path component and a non-line-of-sight (NLoS) path component as
\begin{equation}
    \boldsymbol{H} = \sqrt{\frac{\kappa}{\kappa+1}}\boldsymbol{H}^{\text{LoS}}+\sqrt{\frac{1}{\kappa +1}}\boldsymbol{H}^{\text{NLoS}},
\end{equation}
where $\kappa$ is the Rician factor, while $\boldsymbol{H}^{\text{LoS}}\in \mathbb{C}^{K\times M}$ and $\boldsymbol{H}^{\text{NLoS}} \in \mathbb{C}^{K\times M}$ are the LoS and NLoS components, respectively. We assume that the AAS is a UPA; thus, the LoS channel matrix can be modeled as in \cite[Ch.~7]{bjornson2017massive}.

We model the NLoS channel as $\boldsymbol{h}_k^{\text{NLoS}} \sim \mathcal{CN}(\boldsymbol{0}_M,\boldsymbol{R}_{k})$, which is spatially correlated Rayleigh fading. The spatial correlation matrix ${\boldsymbol{R}_{k}}\in  \mathbb{C}^{M \times M} $ is generated following the local scattering model from \cite[Ch.~2]{bjornson2017massive}. The locations of the UEs are randomly generated with the same elevation angle $\theta =0$ but uniformly distributed azimuth angles $\phi$, seen from the UPA. 
\subsection{Results and Discussion}

We assume the UPA consists of $M = 16$ entries in the form of a $4 \times 4$ array. The number of quantization levels is $L=8$, the number of UEs is $K = 4$, and the Rician factor is  $\kappa = 5$. The UEs have the same SNR, which is defined as $\mathrm{SNR}=\frac{q \gamma}{N_0}$, where $\gamma$ is the common channel variance.

\begin{figure}[!t]
      \centering \vspace{-3mm}
      \includegraphics[scale=0.35]{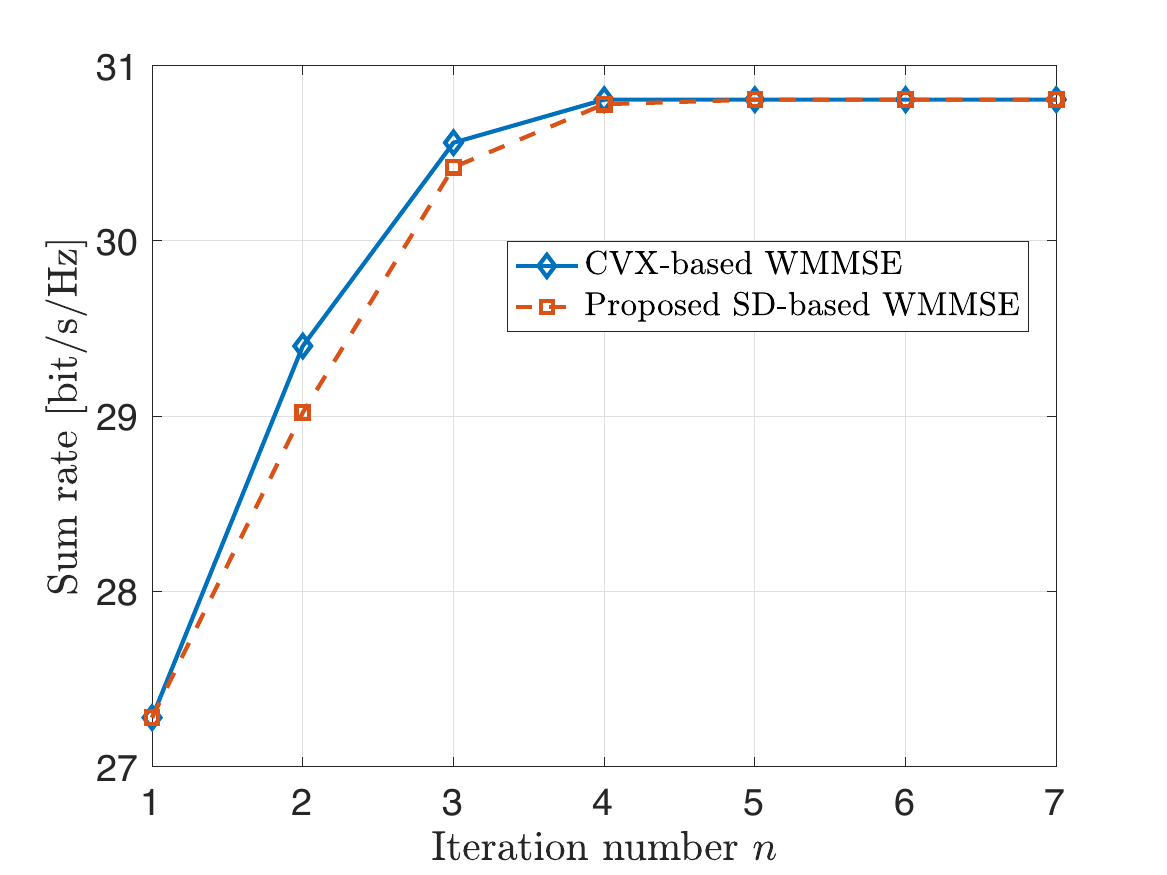}
        \caption{Sum rate evolution when running the proposed WMMSE algorithm using CVX or the proposed SD-based method.}
 \label{fig:iter} \vspace{-3mm}
\end{figure}

Fig.~\ref{fig:iter} presents the convergence behavior of the proposed WMMSE algorithm in Fig.~\ref{fig:algorithm} for the cases when $\mathbb{P}_3$ is solved exactly using CVX (denoted by \emph{CVX-based WMMSE}) and approximately using the proposed SD-based method. We consider $\mathrm{SNR}=20$ dB and generate one user drop. The proposed algorithm reaches a stationary point after $N=5$ iterations. We notice that, despite the approximations made to lower the complexity in the SD-based method, the difference in the sum rate is negligible after convergence. The difference between the sum rate at the starting point and the stationary point approximately shows the improvement of the proposed sum-rate maximization algorithm compared to our previous results in \cite{khorsandmanesh2022optimized}, which considered sum MSE minimization.

\begin{figure}[!t]
        \centering
      \includegraphics[scale=0.35]{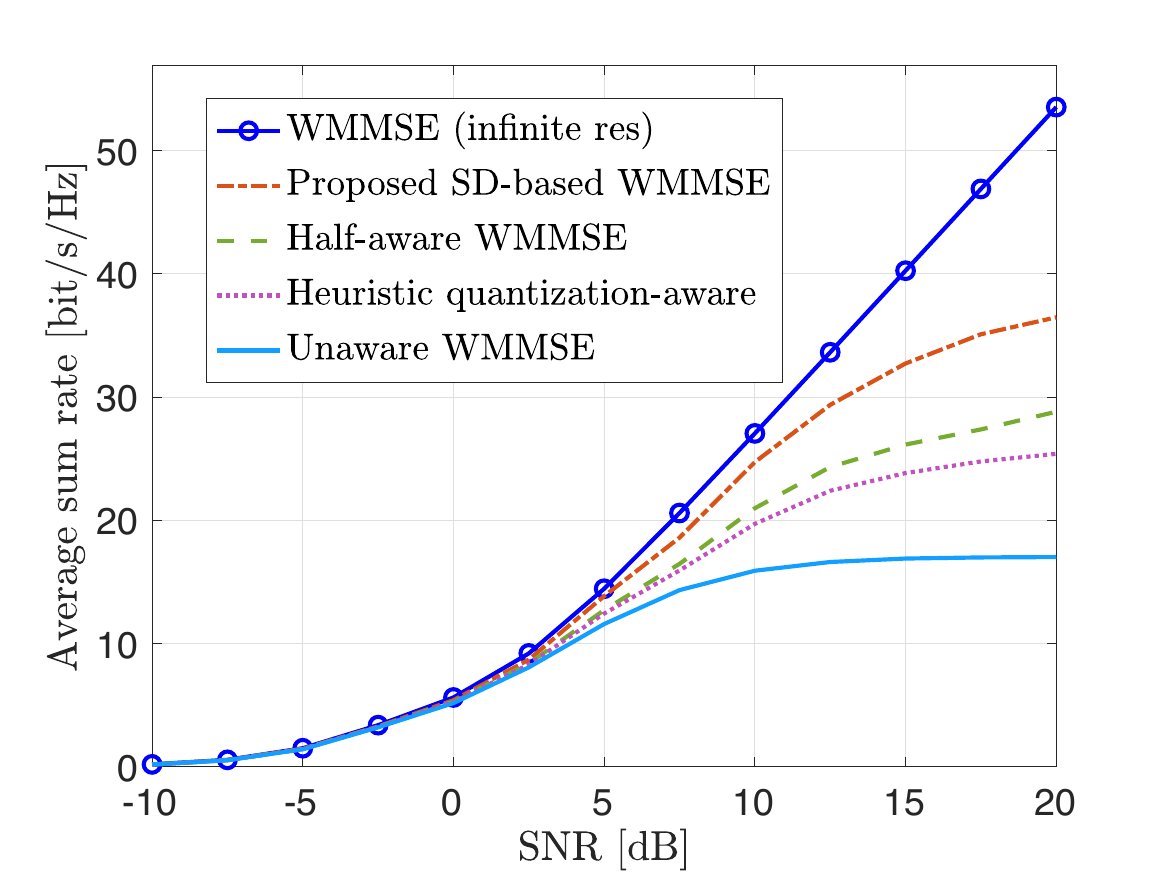}
        \caption{The average sum rate versus the SNR for different precoding schemes. We assume that the BS has $M=4\times4$ antennas and serves $K=4$ UEs with $L=8$ quantization levels.}
 \label{fig:sum} \vspace{-4mm}
\end{figure}

Fig.~\ref{fig:sum} depicts the average sum rate as a function of the SNR for the different precoding schemes. The top curve, \emph{WMMSE (infinite res)}, considers the ideal case without quantization and outperforms all the quantized precoding schemes since the rate increases linearly (in dB scale) at high SNR. In all quantized precoding schemes, the sum rate converges to specific limits at high SNR since the interference cannot be canceled entirely due to the limited precoding resolution; i.e., the system is interference-limited at high SNR. The gap between \emph{WMMSE (infinite res)} and our novel \emph{Proposed SD-based WMMSE} precoding is remarkably smaller than the gap between \emph{Unaware WMMSE} and \emph{WMMSE (infinite res)}, where we quantized the precoding matrix used for \emph{WMMSE (infinite res)}. Our algorithm provides twice the rate at high SNR.
The lower-complexity \emph{Half-aware WMMSE} approach also outperforms the \emph{Unaware WMMSE} algorithm, but there is a substantial gap to \emph{Proposed SD-based WMMSE}. This gap demonstrates the importance of computing quantization-aware weights in the WMMSE algorithm instead of relying on those obtained with infinite resolution. 
The proposed \emph{Heuristic quantization-aware} precoding reaches nearly the same performance as \emph{Half-aware WMMSE} and, thus,  performs vastly better than \emph{Unaware WMMSE} precoding. Note that the complexity of \emph{Heuristic quantization-aware} is polynomial with $M$ and $K$, just as  \emph{Unaware WMMSE}, and therefore is implementable in the same scenarios (e.g., massive MIMO).

\section{Conclusions}

A 5G site often consists of an AAS connected to a BBU via a digital fronthaul with limited capacity. 
Hence, the precoding matrix that is computed at the BBU must be quantized to finite precision. We have proposed a novel WMMSE-based framework for quantization-aware precoding that finds a local optimum to the sum rate maximization problem. Moreover, a reduced-complexity SD algorithm was proposed to find an approximate but tight solution.
We demonstrated that the sum rate can be doubled at high SNR by being quantization-aware both when selecting the weights in the WMMSE formulation and when selecting the precoding matrix.
Besides, a heuristic quantization-aware precoding was proposed to outperform the quantization-unaware baseline while having comparable complexity.
Finally, we note that the framework can be easily modified also to solve  \emph{weighted} sum rate problems.

\bibliographystyle{IEEEtran}
\bibliography{IEEEabrv,references}

\end{document}